\documentclass[
    ,final            
  ]
  {aipproc}

\layoutstyle{6x9}

\usepackage{amssymb}
\usepackage{amsmath}
\usepackage{color}
\newcommand{\pen}{$\pi$$\to$$e$$\nu$\ }
\newcommand{\pme}{$\pi$$\to$$\mu$$\to$$e$\ }
\begin{document}

\title{PEN experiment: a precise measurement of 
                the  $\pi^+\to e^+\nu$ decay branching fraction}

\newcommand{\uva}{Department of Physics, University of Virginia,
                 Charlottesville, VA 22904-4714, USA} 
\newcommand{\dubna}{Joint Institute for Nuclear Research, RU-141980 Dubna,
                 Russia} 
\newcommand{\PSI}{Paul Scherrer Institut, Villigen PSI, CH-5232, Switzerland}
\newcommand{\IRB}{Institut ``Rudjer Bo\v{s}kovi\'c,'' HR-10000 Zagreb,
                 Croatia} 
\newcommand{\swierk}{Instytut Problem\'ow J\k{a}drowych im.\ Andrzeja
                 So{\l}tana, PL-05-400 \'Swierk, Poland} 
\newcommand{\tbilisi}{Institute for High Energy Physics, Tbilisi State
                 University, GUS-380086 Tbilisi, Georgia} 
\newcommand{\unizh}{Physik-Institut, Universit\"at Z\"urich, CH-8057
                 Z\"urich, Switzerland}

\author{D.~Po\v{c}ani\'c}{address=\uva}
\author{L.~P.~Alonzi}{address=\uva}
\author{V.~A.~Baranov}{address=\dubna}
\author{W.~Bertl}{address=\PSI}
\author{M.~Bychkov}{address=\uva}
\author{Yu.M.~Bystritsky}{address=\dubna}
\author{E.~Frle\v{z}}{address=\uva}
\author{V.A.~Kalinnikov}{address=\dubna}
\author{N.V.~Khomutov}{address=\dubna}
\author{A.S.~Korenchenko}{address=\dubna}
\author{S.M.~Korenchenko}{address=\dubna}
\author{M.~Korolija}{address=\IRB}
\author{T.~Kozlowski}{address=\swierk}
\author{N.P.~Kravchuk}{address=\dubna}
\author{N.A.~Kuchinsky}{address=\dubna}
\author{D.~Mekterovi\'c}{address=\IRB}
\author{D.~Mzhavia}{address=\dubna,altaddress=\tbilisi}
\author{A.~Palladino}{address=\uva,altaddress=\PSI}
\author{P.~Robmann}{address=\unizh}
\author{A.M.~Rozhdestvensky}{address=\dubna}
\author{S.N.~Shkarovskiy}{address=\dubna}
\author{U.~Straumann}{address=\unizh}
\author{I.~Supek}{address=\IRB}
\author{P.~Tru\"ol}{address=\unizh}
\author{Z.~Tsamalaidze}{address=\tbilisi}
\author{A.~van~der~Schaaf}{address=\unizh}
\author{E.P.~Velicheva}{address=\dubna}
\author{V.P.~Volnykh}{address=\dubna}

\begin{abstract}
A new measurement of $B_{\pi e2}$, the $\pi^+ \to e^+\nu(\gamma)$
decay branching ratio, is currently under way at the Paul Scherrer
Institute. 
The present experimental result on $B_{\pi e2}$ constitutes the most
accurate test of lepton universality available. The accuracy, however,
still lags behind the theoretical precision by over an order of
magnitude.
Because of the large helicity suppression of the $\pi_{e2}$ decay, its
branching ratio is susceptible to significant contributions from new
physics, making this decay a particularly suitable subject of study.

\end{abstract}

\maketitle


\section{Motivation}
\label{sec:motiv}

Historically, the $\pi$$\to$$e\nu$ decay, also known as $\pi_{e2}$,
provided an early strong confirmation of the $V-A$ nature of the
electroweak interaction.  At present, thanks to exceptionally well
controlled theoretical uncertainties, its branching ratio is
understood at the level of better than one part in $10^4$.  The most
recent independent theoretical calculations are in very good agreement
and give:
\begin{equation}
     B_{\text{calc}}^{\text{SM}} = 
       \frac{\Gamma(\pi \to  e\bar{\nu}(\gamma))}
          {\Gamma(\pi \to  \mu\bar{\nu}(\gamma))}\bigg|_{\text{calc}} =
 \begin{cases}
    1.2352(5) \times 10^{-4} & \text{Ref.~\cite{Mar93},} \\
    1.2354(2) \times 10^{-4} & \text{Ref.~\cite{Fin96},} \\
    1.2352(1) \times 10^{-4} & \text{Ref.~\cite{Cir07},} \\
   \end{cases}
\end{equation}
where $(\gamma)$ indicates that radiative decays are included.
Marciano and Sirlin \cite{Mar93} and Finkemeier \cite{Fin96} took into
account radiative corrections, higher order electroweak leading
logarithms, short-distance QCD corrections, and structure-dependent
effects, while Cirigliano and Rosell \cite{Cir07} used the two-loop
chiral perturbation theory.  A number of exotic processes outside of
the current standard model (SM) can produce deviations from the above
predictions, mainly through induced pseudoscalar (PS) currents.  Prime
examples are: charged Higgs in theories with richer Higgs sector than
the SM, PS leptoquarks in theories with dynamical symmetry breaking,
certain classes of vector leptoquarks, loop diagrams involving certain
SUSY partner particles, as well as non-zero neutrino masses and their
mixing (Ref.~\cite{Rai08} gives a recent review of the subject).  In
this sense, $\pi_{e2}$ decay provides complementary information to
direct searches for new physics at modern colliders.

The two most recent measurements of the branching ratio are mutually
consistent:
\begin{equation}
  B_{\text{exp}}^{\pi e2} =
   \begin{cases}
      1.2265\,(34)_{\text{stat}}\,(44)_{\text{syst}}
            \times 10^{-4} & \text{Ref.~\cite{Bri92},}  \\
      1.2346\,(35)_{\text{stat}}\,(36)_{\text{syst}}
            \times 10^{-4} & \text{Ref.~\cite{Cza93},}  \\
   \end{cases}
\end{equation}
and dominate the world average of $1.2300\,(40) \times 10^{-4}$,
which, however, is less accurate than the theoretical calculations by
a factor of 40.  The PEN experiment \cite{pen06} is aiming to reduce
this gap and, in doing so, to set new limits on the above non-SM
processes.

\section{Experimental method} 
\label{sec:method}

\begin{figure}[tb]
  \parbox{0.57\linewidth}{
     \includegraphics[width=\linewidth]{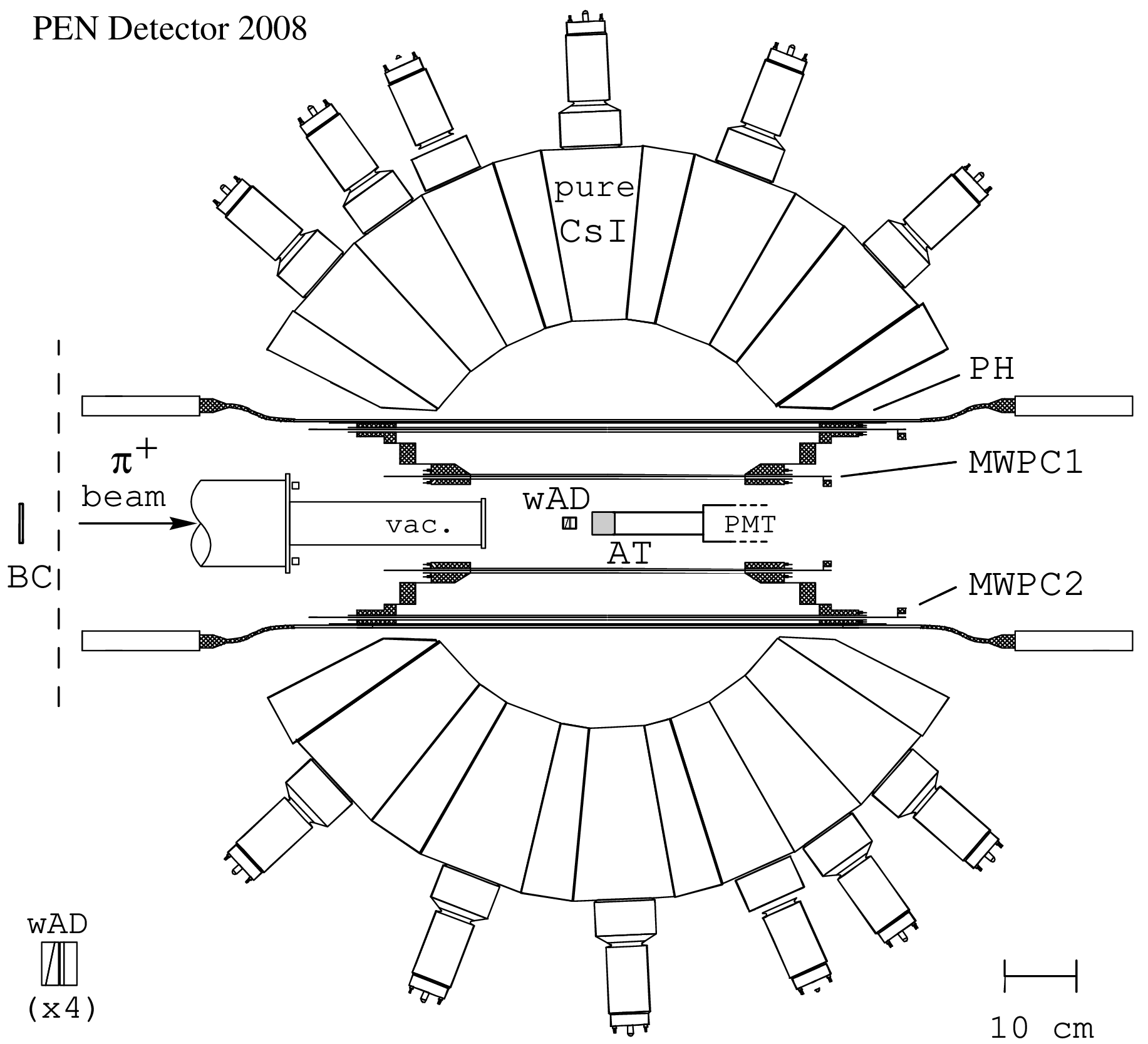} 
                         }
  \parbox{0.42\linewidth} {
     \includegraphics[width=\linewidth]{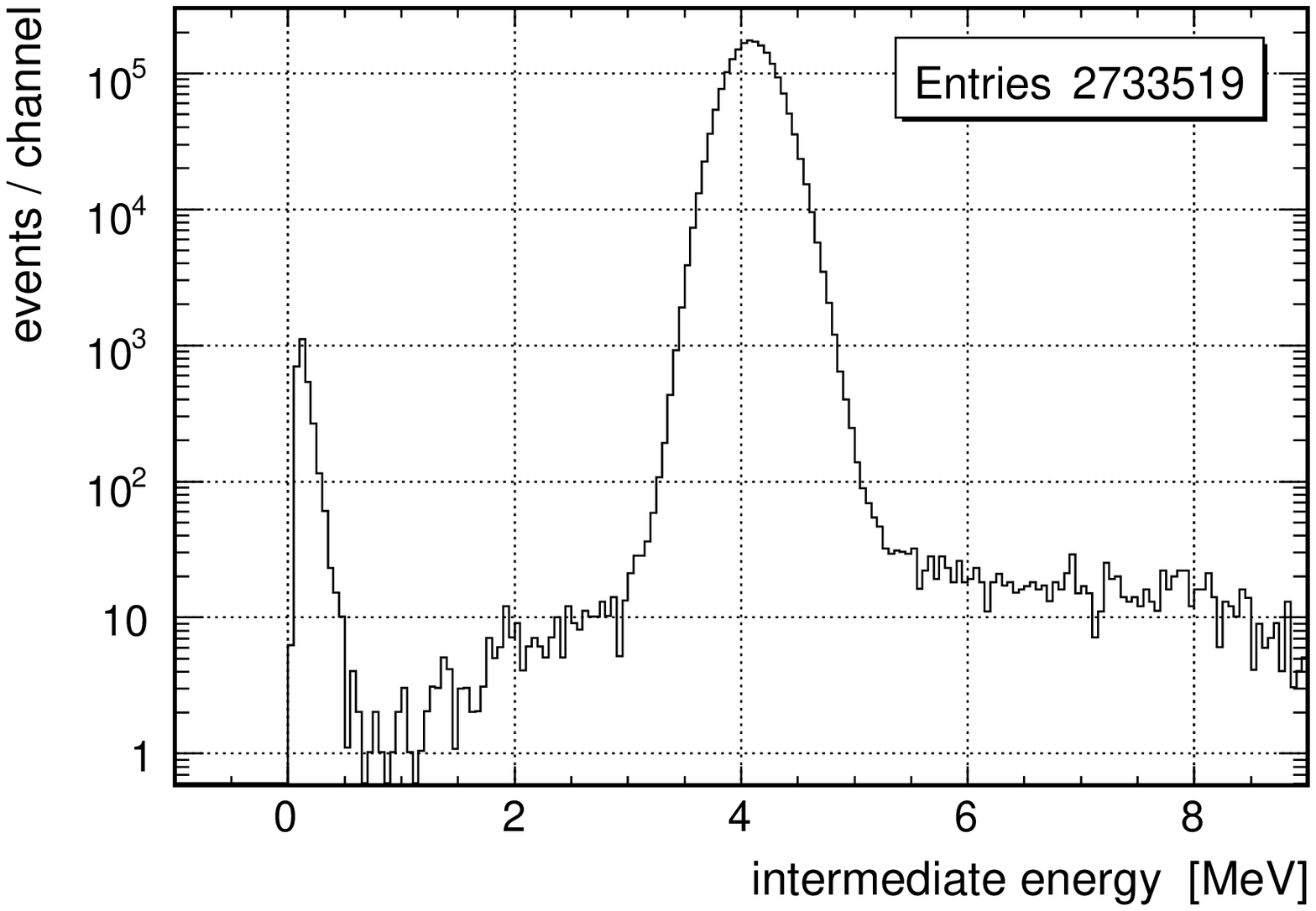}
     \includegraphics[width=\linewidth]{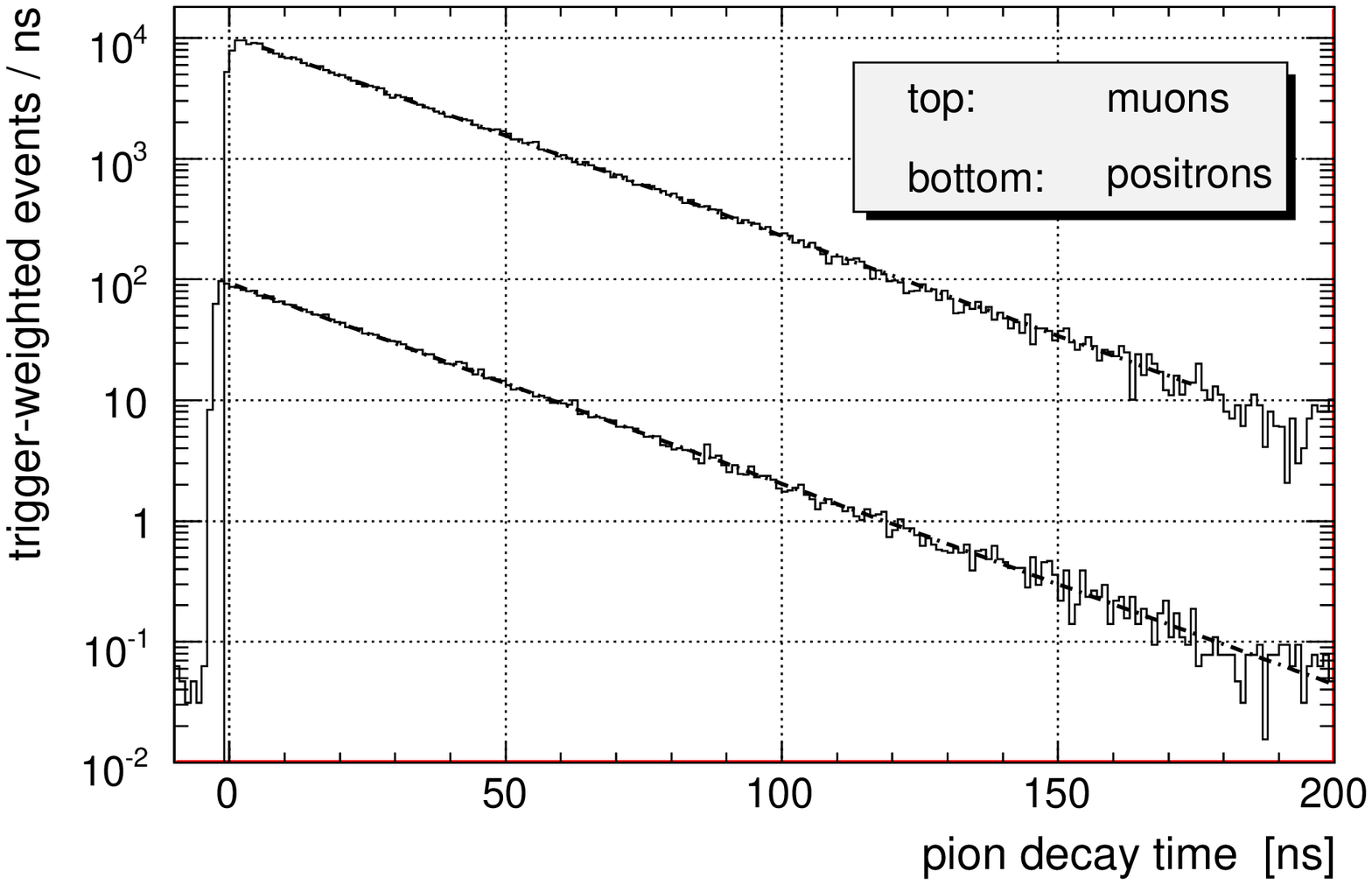}
                          }
    \caption{Left: Cross section drawing of the PEN detector system.
      A small schematic representation of the wedged Active Degrader
      (wAD) is included in the lower left corner.  
      Right: Energy distribution of the intermediate muon in the \pme
      decay chain,  extracted from AT waveform data (top).  Time
      dependence of the $\pi$$\to$$\mu$ and $\pi$$\to$$e$ decays,
      from the AT waveform data.
      \label{fig:pen_det_res} }
\end{figure}

The PEN experiment uses an upgraded version of the PIBETA detector
system, described in detail in Ref.~\cite{Frl04a}.  The PIBETA
collaboration performed a series of rare pion and muon decay
measurements \cite{Poc04,Frl04b,Byc08}.  The PEN apparatus consists of
a large-acceptance ($\sim 3\pi$\,sr) electromagnetic shower
calorimeter (pure CsI, 12 radiation lengths thick) with non-magnetic
tracking in concentric cylindrical multi-wire proportional chambers
(MWPC1,2) and plastic scintillator hodoscope (PH), surrounding a
plastic scintillator active target (AT).  Beam pions pass through an
upstream detector (BC), have their energy reduced in the wedged active
degrader (wAD), and stop in the target.  Combining the signals from
the four wedges (a horizontal pair and a vertical pair) of the wAD
provides information on the $x$-$y$ position of the beam pion as it
reaches the AT.  Signals from the beam detectors (BC, wAD, and AT) are
digitized in a 2\,GS/s waveform digitizer.  Figure
\ref{fig:pen_det_res} shows the layout of the main detector
components.

\begin{figure}[b]
  \includegraphics[width=0.45\linewidth]{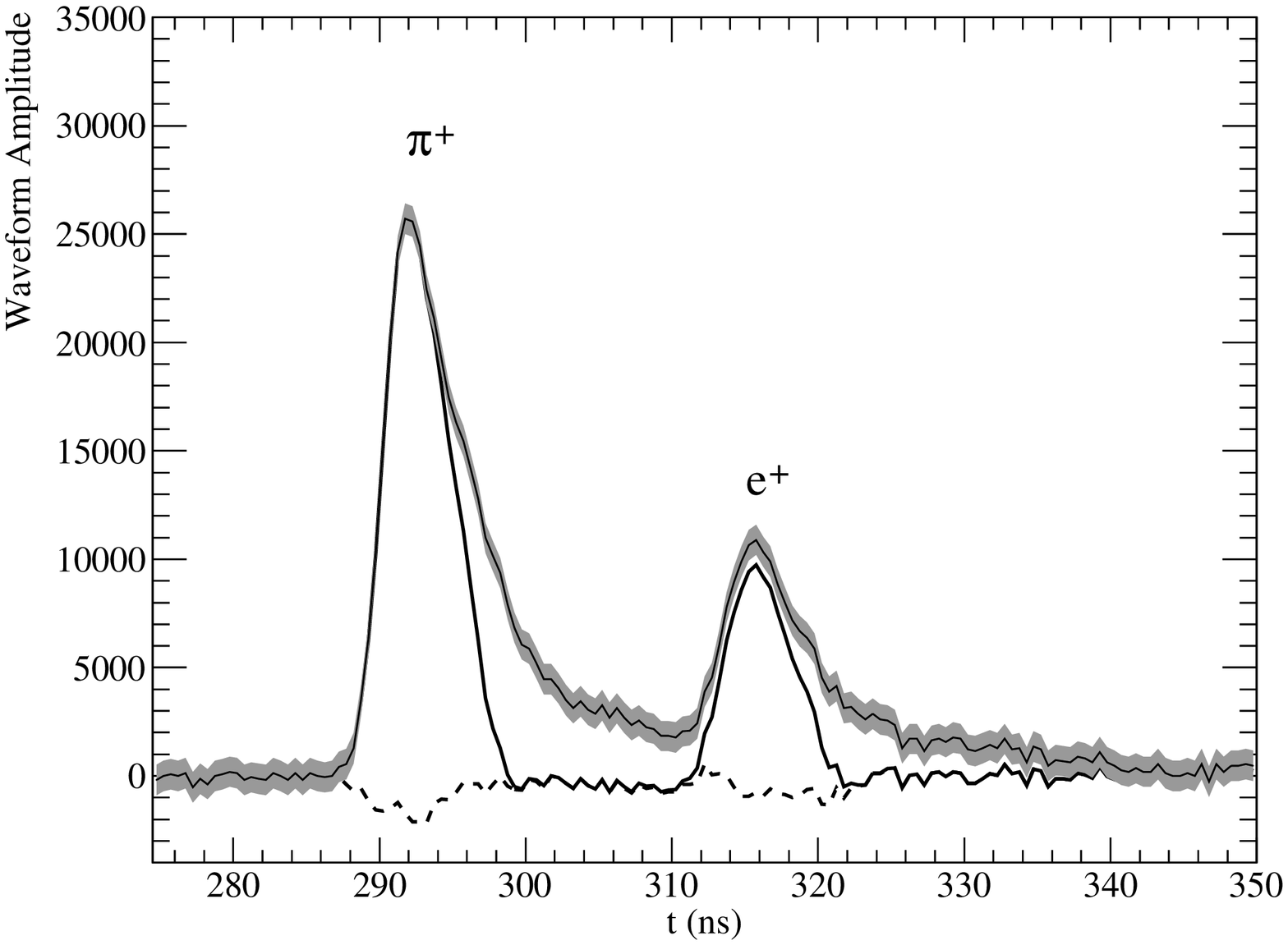}
  \hspace*{\fill}
  \includegraphics[width=0.45\linewidth]{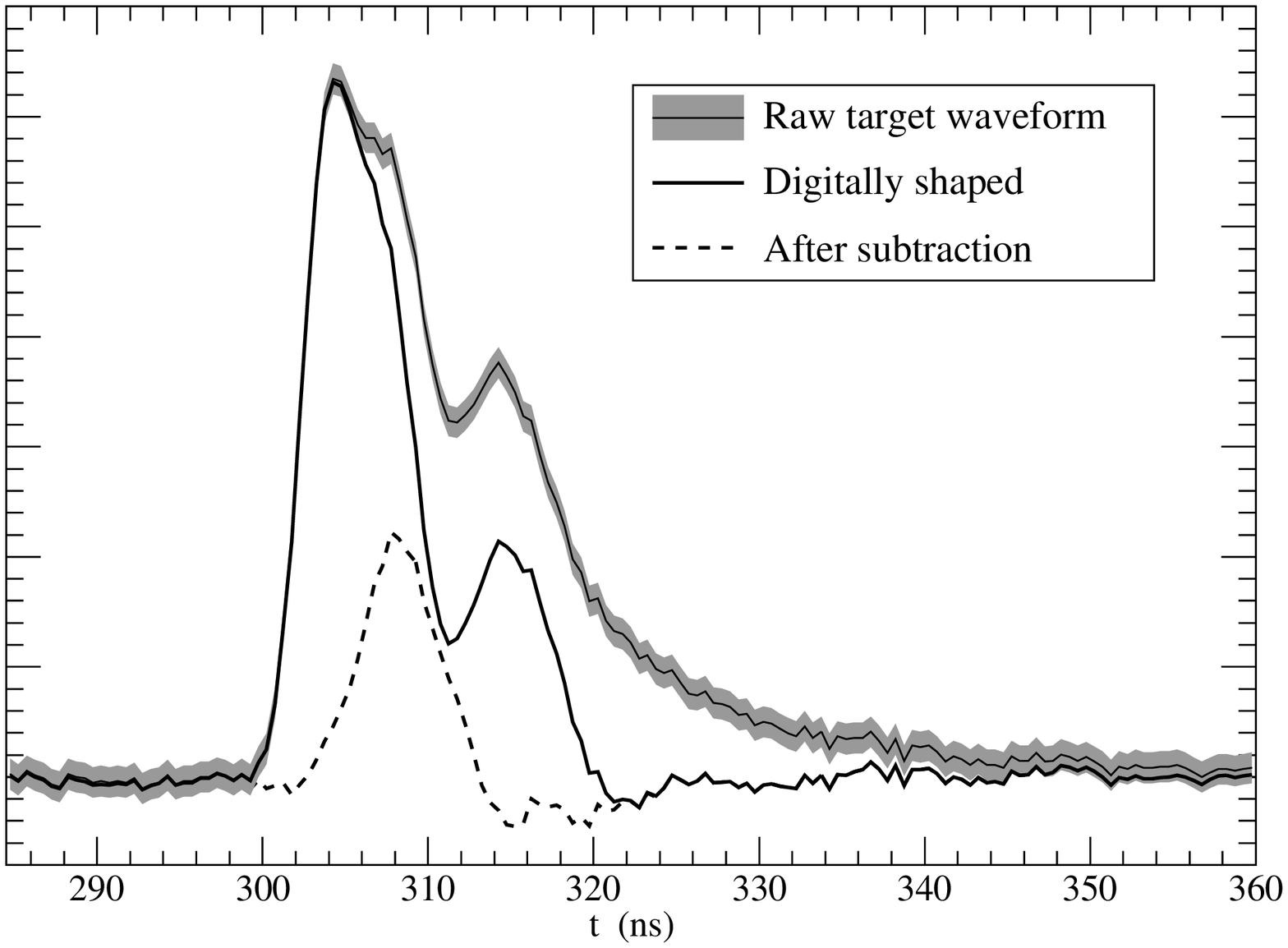}
    \caption{Target detector waveforms: raw (solid with shaded
      uncertainty band), and digitally shaped (solid curve) to
      suppress the pulse tails.  Left: an event in which no residual
      pulse (dashed curve) was found after subtracting the stopped
      $\pi^+$ and emitted e$^+$ pulses.  Right: an event in which a
      closely spaced intermediate muon pulse (dashed curve) was found
      after subtracting the $\pi^+$ and e$^+$ pulses.  Uncertainty
      bands for the shaped and subtracted waveforms (not shown) are of
      similar size to the raw ones.
      \label{fig:tgt_wavfrms} }
\end{figure}

\begin{figure}[tb]
  \parbox{0.49\linewidth}{
     \includegraphics[width=\linewidth]{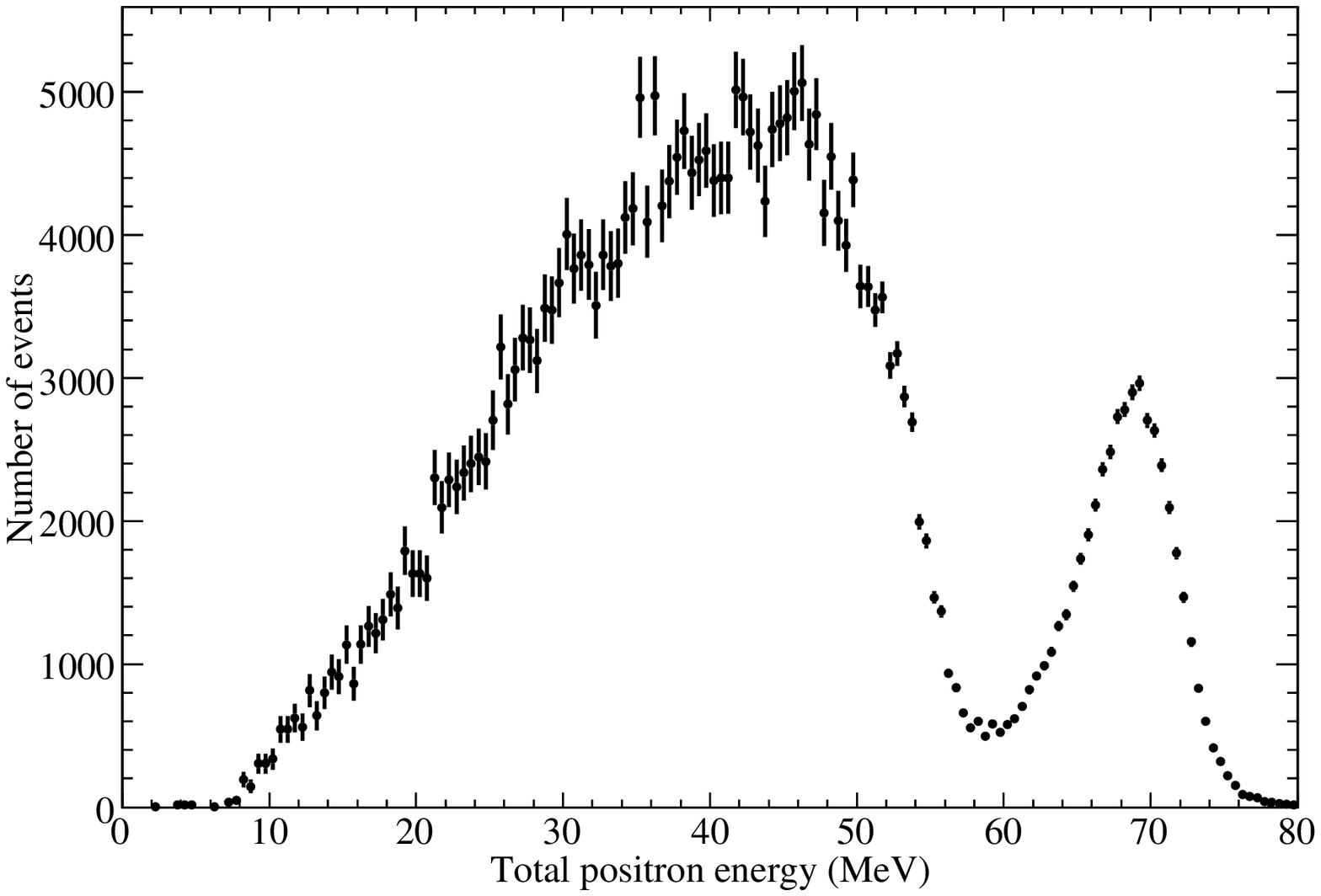} 
                         }
  \hspace*{\fill}
  \parbox{0.49\linewidth} {
     \includegraphics[width=\linewidth]{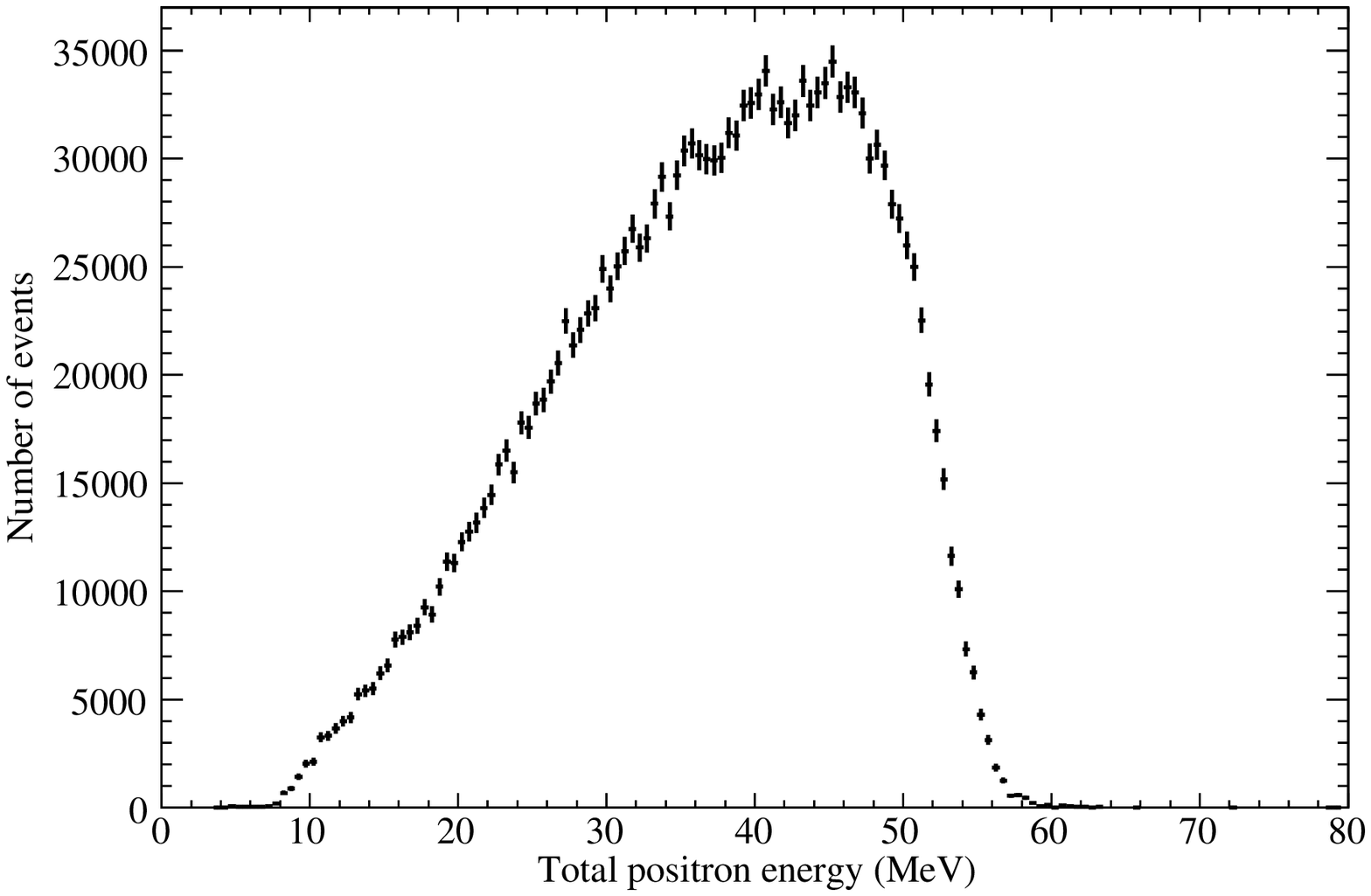}
                          }
    \caption{Measured $e^+$ energy spectrum for early (0 to 50\,ns;
      left), and late decay times (200 to 250\,ns; right).  Early
      decays show a distinct \pen peak above the \pme continuum
      (additionally suppressed by a target energy cut).  The \pen peak
      is essentially gone by 200\,ns.
      \label{fig:epls_spectra} }
\end{figure}

The primary method of evaluating the $\pi_{e2}$ branching ratio, as
outlined in the experiment proposal \cite{pen06}, is to normalize the
observed yield of \pen decays in a high-threshold calorimeter energy
trigger (HT) to the number of sequential decays \pme in an
all-inclusive pre-scaled trigger (PT).  To be accepted in either
trigger, an event must contain the positron signal within a 250\,ns
gate which starts about 40\,ns before the pion stop time.  The
$\pi_{e2}$ events are isolated and counted within the HT data sample
via their 26\,ns exponential decay time distribution with respect to
the pion-stop time reference.  On the other hand, the PT provides the
corresponding number of sequential \pme decays, again via their well
defined time distribution with respect to the $\pi$ stop time.
Finally, by analyzing the beam counter waveform digitizer data we
separate $\pi_{e2}$ events (two pulses in the target waveform: pion
stop and decay positron), from the sequential decay events (three
pulses in the target waveform, produced by the pion, muon and
positron, respectively).  Thus identified $\pi_{e2}$ events serve to
map out the energy response function of the calorimeter, enabling us
to evaluate the low energy ``tail,'' not accessible in the HT data.  A
measure of the performance of the beam detectors, and of the current
state of the art of the waveform analysis is given in
Fig.~\ref{fig:pen_det_res} (right), showing excellent identification
of the \pme and \pen decays.  The underlying waveform analysis is
illustrated in Fig.~\ref{fig:tgt_wavfrms}.
Figure \ref{fig:epls_spectra} shows the positron energy spectra for
$\pi_{e2}$ decays and $\mu\to e$ decays recorded in our detector.

Two engineering runs were completed, in 2007 and 2008, respectively.
All detector systems are performing to specifications, and $\sim 5
\times 10^6$ $\pi_{e2}$ events were collected.  The experiment will
continue with a major run in 2009, which is intended to double the
existing data sample.  Prior to the run, the wedged degrader will be
replaced by a single thin degrader, and a mini time projection chamber
(mini-TPC).  The position resolution, $\sim$1--2\,mm, achieved using
the wAD, is limited by pion multiple scattering in the wedges.  The
new system will improve the beam position resolution by about an order
of magnitude, with the mini-TPC adding excellent directional
resolution, as well.  Both are needed to achieve improved systematics
and better control of decays in flight.

%
%

\begin{theacknowledgments}
  This work has been supported by grants from the US National Science
  Foundation (most recently PHY-0653356), the Paul Scherrer Institute,
  and the Russian Foundation for Basic Research (Grant 08-02-00652a).
\end{theacknowledgments}

\end{document}